\documentclass{article}
\usepackage{amssymb}
\usepackage{amsmath}

\begin{document}
\large
 \baselineskip18pt

\title{Random Dynamics, Entropy Production and  Fisher Information}
\author{Piotr Garbaczewski\thanks{Presented at the XV Marian Smoluchowski
Symposium on Statistical  Physics, Zakopane, Poland, September 7-12, 2002} \\
Institute of Physics,  University  of Zielona  G\'{o}ra,
65-516 Zielona G\'{o}ra, Poland}
\maketitle
\begin{abstract}
We analyze a  specific role  of probability density gradients  in the theory of
irreversible transport processes.  The classic Fisher information and information
 entropy production concepts are found to be intrinsically entangled with
 the very notion of  the Markovian diffusion process and that of the related (local)
 momentum conservation law.
\end{abstract}

\section{Motivations and  associations}

The main objective of the present paper is to analyze the role -
origins, possible  physical  meaning and manifestations - of \it
two \rm analytical  expressions which are  omnipresent, directly
or indirectly, in  any  theoretical framework addressing  an issue
of transport  driven by Markovian diffusion processes.  Both
derive from the sole  properties, and specifically the time
evolution,  of the probability density associated  with the
analyzed  stochastic process (like e.g. the dynamics of tracer
particles in  a gas or fluid).

Let us specify the context by considering spatial Markov diffusion
processes with a diffusion parameter (constant or time-dependent)
$D$ and generally  space-time inhomogeneous probability density $\rho $.

 \it One  \rm of the aforementioned expressions  reads:
\begin{equation}
Q=2D^{2}{\frac{\Delta \rho ^{1/2}} {\rho ^{1/2}}} =
D^2 \left[ {\frac{1}\rho }\Delta \rho -
{\frac{1}2}{\frac{1}{\rho ^2}} (\overrightarrow{\nabla }\rho )^2 \right] =
{\frac{1}2}\overrightarrow{u}^2 + D\overrightarrow{\nabla} \cdot \overrightarrow{u}
\end{equation}
where  $\overrightarrow{u} = D \overrightarrow{\nabla } \ln \rho $ is sometimes named
an osmotic velocity field.

Density gradients are here explicitly involved and  it is useful to invoke at this point
 a vivid discussion, carried out  recently,  about the status of density gradient
 as a "real" (thermodynamic) force performing work on the particles and the related issue
 of the  local irreversible entropy production, see  e.g.
 \cite{cohen}-\cite{gaspard2}, see however also \cite{gar}-\cite{czopnik}.

 Let us recall that the standard spatial Brownian motion involves
 $\overrightarrow{v} = - \overrightarrow{u}$, known  as the
diffusion current velocity and  (up to a dimensional factor)
identified with the "thermodynamic force  of diffusion"
\cite{gaspard2} which drives  the irreversible process of matter
exchange  at the macroscopic level. In terms of tracer particles,
this irreversible process occurs even if they are so dilute that
they never meet (nor interact with) each other.

On the other hand, even while the "thermodynamic force" is  a
concept  of purely statistical origin associated  with a
collection of particles, in contrast to microscopic  forces which
have a direct impact on individual particles themselves, it is
well known  \cite{gar,gar1,gaspard2}  that this force manifests
itself as a Newtonian-type entry   in  local conservation laws
describing the momentum balance: in fact that pertains to  the
average (local averages) momentum taken over by the "particle
cloud", a statistical ensemble   property quantified in terms of
the probability distribution at hand.
It is precisely  the (negative) gradient of the above  potential  $Q$,
Eq. (1), which  plays the Newtonian force role in the momentum balance
equations, \cite{gar,gar1}.

To elucidate the role of the  \it second \rm  analytical
expression of interest  in our present considerations ($Q$ was actually
the \it first) \rm  let us observe that in  one space dimension, for probability
densities vanishing at spatial infinities,  we have:
\begin{equation}
- \int Q \rho dx = \int
{\frac{u^2}2} \rho dx \doteq {\frac{1}2} D^2 \cdot F_X
\end{equation}
where  $F_X$ is the
so-called Fisher information \cite{frieden,luo,carlen}
 of (encoded in)  the probability density
$\rho $ which quantifies its "gradient content"
(sharpness plus localization/disorder properties) and reads:
\begin{equation}
F_X = \int {\frac{(\nabla \rho )^2}{\rho }} dx \, .
\end{equation}

An important property of the \it Fisher information \rm  (stemming
from the Cramer-Rao inequality in the statistical inference
theory, \cite{frieden}-\cite{cramer}) is that $F_X^{-1}$ sets
the lower bound for the variance of the random variable $X(t)$ with
values in $R^1$, distributed according to $\rho (x,t)$:
\begin{equation}
\left<X^2 \right> - \left< X \right>^2 \geq F^{-1}_X \, .
\end{equation}

On the other hand, in direct correspondence with our previous discussion of
$Q(x,t)$, let us point out that  the integrand in Eq.(3),  up to a dimensional
factor, defines the so-called \it local entropy production \rm inside the
system sustaining an irreversible process of diffusion, \cite{cohen,gaspard}.
Accordingly, \cite{gaspard,gaspard1}
\begin{equation}
{\frac{dS}{dt}} = D\cdot \int {\frac{(\nabla \rho )^2}{\rho }} dx  = D \cdot F_X \geq 0
\end{equation}
stands for an entropy production rate   when the Fick law-induced
diffusion current (standard Brownian motion case)   $j =
- D \nabla \rho $,  obeying $\partial _t \rho + \nabla j = 0$,
enters the scene. Here $S= - \int \rho \ln \rho \,  dx$ plays the
role of the (time-dependent) \it  information  entropy \rm in
the nonequilibrium statistical mechanics framework  for the thermodynamics
of irreversible processes. It is rather clear that the high rate of the  entropy increase
corresponds to a rapid spreading (flattening down) of the probability density.
That explicitly  depends on the "sharpness " of density gradients.

The potential-type function $Q(x,t)$,  the Fisher information
$F_X(t)$, nonequilibrium measure of the entropy production
$dS/dt$ and the information  entropy   $S(t)$ are thus
mutually  entangled quantities, each  being exclusively
determined in terms of the probability density $\rho (x,t)$ and
its spatial derivatives.

\section{Hydrodynamical (local) momentum conservation laws - the zoo}

As mentioned before, the function $Q(x,t)$  notoriously appears
in various local conservation laws responsible for the momentum
balance in suitable physical systems. Let us make a brief perusal
of the respective partial differential equations.

In the standard  statistical mechanics  setting, the  Euler
equation does not refer to any $Q$, Eq. (1), but  deserves reproduction
for the  obvious comparison purpose as a prototype momentum balance
equation in the (local) mean:
\begin{equation}
 (\partial _t + \overrightarrow{v} \cdot
\overrightarrow{\nabla }) \overrightarrow{v}  =
\frac{\overrightarrow{F}}{\ m}\,  -  \,
\frac{\overrightarrow{\nabla } P} \rho
 \end{equation}
where we generally assume  $\overrightarrow{F} = - \overrightarrow{\nabla } V$ to represent
the  "normal" Newtonian force.

With regard to the  manifest appearance  of $Q$, we  begin from an
encounter  with $\overrightarrow{\nabla }Q$ in an
out-of-statistical mechanics example provided by the
hydrodynamical formalism of quantum theory, \cite{holland}:
\begin{equation}
 (\partial _t + \overrightarrow{v} \cdot
\overrightarrow{\nabla }) \overrightarrow{v} = {\frac{1}m}
\overrightarrow{F} - \overrightarrow{\nabla } Q_q = {\frac{1}m}
\overrightarrow{F} + {\frac{\hbar
^2}{2m^2}} \overrightarrow{\nabla }
{\frac{\Delta \rho ^{1/2}}{\rho ^{1/2}}}
\end{equation}
 where $Q_q= - {\frac{\hbar
^2}{2m^2}}{\frac{\Delta \rho ^{1/2}}{\rho ^{1/2}}}$  is the
familiar  de Broglie - Bohm quantum potential.

Another  spectacular example  pertains to the standard free
Brownian motion in the strong friction (Smoluchowski diffusion)
regime. Namely, we have, \cite{gar}:
\begin{equation}
(\partial _t + \overrightarrow{v}\cdot \overrightarrow{\nabla
})\overrightarrow{v} = - 2D^2 \overrightarrow{\nabla }
\frac{{\Delta  }\rho ^{1/2}}{\rho ^{1/2}} \doteq - \overrightarrow{\nabla } Q
\end{equation}
 where $\overrightarrow{v}= -  D \frac{\overrightarrow{\nabla }\rho }
\rho $;  $D$ is the diffusion constant  (set formally $D\doteq
\hbar/2m$ and  notice the sign change in comparison  with the
previous quantum mechanical law).

The large friction  (Smoluchowski again) limit of the driven
phase-space random dynamics implies, \cite{gar1}:
\begin{eqnarray}
( \partial _{t} + \overrightarrow{v}
\cdot \overrightarrow{\nabla }) \overrightarrow{v}%
 &=& \overrightarrow{\nabla }
\left( \Omega -Q\right)
\end{eqnarray}
where $\overrightarrow{v} \doteq
\overrightarrow{v}(\overrightarrow{x},t) = \frac{%
\overrightarrow{F}}{\ m\beta }- D\frac{\overrightarrow{\nabla }
\rho }{\rho }$, the volume force (notice the positive
sign)  reads $ + \overrightarrow{\nabla }\Omega $ instead of
the previous $-\overrightarrow{\nabla }V$.
Here $Q=2D^{2}\frac{\Delta \rho ^{1/2}}{\rho ^{1/2}}$ and  (recall
the spectral analysis of Fokker-Planck operators, cf. \cite{gar1})
\begin{equation}
\Omega =\frac{1}{2}\left( \frac{\overrightarrow{F}}{\ m\beta
}\right) ^{2}+D\overrightarrow{\nabla }\cdot \left(
\frac{\overrightarrow{F}}{\ m\beta }\right)
\end{equation}

For a class of "perverse" diffusion processes (respecting the
so-called "Brownian recoil principle", \cite{gar1}), we deal with
Markovian diffusion processes with the  inverted sign of
$\overrightarrow{\nabla }(\Omega - Q)$ in the  local momentum
conservation law, so that the  previous Eq. (9)  takes the form:
\begin{eqnarray}
( \partial _{t} + \overrightarrow{v}
\cdot \overrightarrow{\nabla }) \overrightarrow{v}%
 &=& \overrightarrow{\nabla }
\left( Q -\Omega \right)
\end{eqnarray}
By  introducing  ${\psi =
\rho ^{1/2} exp(iS)}$ and  $\overrightarrow{v} = 2D
\overrightarrow{\nabla } S$, we set  a  link with
the "true" (notice an imaginary unit $i$) Schr\"{o}dinger-type dynamics:
\begin{equation*}
i\partial _t \psi = - D\Delta \psi + {\frac{\Omega }{2mD}}\, \psi
\end{equation*}
Useful observation: the total energy
$\int_{R^3}({{\overrightarrow{v}}^2\over 2} -Q + \Omega )\rho
d^3x$ = $\int_{R^3}({{\overrightarrow{v}}^2\over 2} +
{{\overrightarrow{u}}^2\over 2} + \Omega )\rho d^3x$ of the
system  is a  conserved finite quantity. Here
$\overrightarrow{u}(\overrightarrow{x},t) \doteq D
\overrightarrow{\nabla }\ln \rho (\overrightarrow{x},t)$. Notice
that $(D/2) dS/dt$ of  Eq.(5), makes  an  explicit contribution
to an overall energy of the system.

The conservation of the  total
energy tells that the entropy production and the kinetic energy
due to diffusion currents stay in competition.

For a special case of the   frictionless random phase-space dynamics,
\cite{czopnik}, we arrive at:
\begin{equation}
\left[ \partial _{t}+ \overrightarrow{v} \cdot \overrightarrow{\nabla }\right]
\overrightarrow{v} = \frac{\overrightarrow{F}}m
+2d^2(t)
\overrightarrow{\nabla }\left[ \frac{%
\Delta \rho ^{1/2}}{\rho ^{1/2}}\right]
\end{equation}
where $\overrightarrow{F}$ denotes  the external force  acting on the
particle, and $d(t)$ is the time-dependent diffusion parameter.
This form of the law has been derived by explicitly solving the
Fokker-Kramers equation with properly adjusted (gaussian densities)
 initial data,  for the following cases:
\begin{enumerate}
\item  free particle:  $F\equiv 0$, $n=1$

\item  charged particle in a constant magnetic field:
$\overrightarrow{F}= e \overrightarrow{v} \times \overrightarrow{B}$, $n=2$

\item  harmonically bound particle:  $F=- m \omega ^{2}x$, $n=1$

\end{enumerate}

Presumably this form is  universal (no general  proof at the moment).
The coefficient $d^2(t)$ in all those cases can  be represented as a product
of variances (n=1) evaluated with respect to  conditioned phase-space
$ w_x(u,t) = \frac{f\left( x,u,t\right) }{\rho \left( x,t\right) }$
 and configuration space (marginal)   $\rho \left( x,t\right) = \int f\left( x,u,t\right) du$
 densities respectively:
\begin{equation}
d^{2}\left( t\right) \doteq \left( \left\langle u^{2}\right\rangle
_{x}-\left\langle u\right\rangle _{x}^{2}\right) \left( \left\langle
x^{2}\right\rangle -\left\langle x\right\rangle ^{2}\right)
\end{equation}
One may prove  that $d^2(t)$ is bounded from below which results
in the Heisenberg-type inequality for variances: of $U(t)$ with
respect to the conditioned phase-space density $w_x(u,t)$, and $X(t)$ with
respect to the marginal density $\rho (x,t)$.

\section{Diffusion processes and differential  equations - pedestrian
reasoning}

Let us sketch how the previous observations come out within the
traditional setting of phase-space stochastic processes.

\subsection{Standard Brownian  motion}

Let us consider the  competition between deterministic/random
driving and friction in the standard Brownian motion:
\begin{eqnarray}
\frac{d\overrightarrow{x}}{dt} &=&\overrightarrow{u} \\
\frac{d\overrightarrow{u}}{dt} &=&-\beta \overrightarrow{u}+\frac{%
\overrightarrow{F}}{m}+\overrightarrow{A}\left( t\right)
\end{eqnarray}
where
 $\left\langle A_{i}\left( s\right) \right\rangle =$ $0$ and
 $\left\langle A_{i}\left( s\right) A_{j}\left( s^{\shortmid }\right)
\right\rangle =2q\delta \left( s-s^{\shortmid }\right) \delta
_{ij}$; $\overrightarrow{F} = - \overrightarrow{\nabla } V$.

For the case of the  \it standard \rm Brownian motion,  we know a
priori, in view of the fluctuation-dissipation theorem, that
$q=D\beta ^{2}$  where  $D=\frac{kT}{m\beta }$, while $\beta $ is
given  by the Stokes formula $m\beta =6\pi \eta a$.

The resulting (Markov)  phase - space diffusion  process is
determined by solutions of the Kramers equation: an initially
given $f(\overrightarrow{x}_0,\overrightarrow{u}_0,t_0)$ is
propagated according to:
\begin{equation}
\left( \partial _t + \overrightarrow{u}\cdot
\overrightarrow{\nabla }_{\overrightarrow{x}} +
{\frac{\overrightarrow{F}}{m}}\cdot  \overrightarrow{\nabla }
_{\overrightarrow{u}}
 \right)
f = C(f) =  \left( q\nabla _{\overrightarrow{u}}^{2}  + \beta
\overrightarrow{u} \cdot \overrightarrow{\nabla
}_{\overrightarrow{u}} \right) f
\end{equation}

 Here we adopt the kinetic theory notation for a  substitute of collision
 term, where  $\int C(f) d^3u = 0$,  while $ {\frac{1}\rho }\int
\overrightarrow{u} C(f) d^3u =
- \beta \overrightarrow{v}(\overrightarrow{x},t)$.

Accordingly, the continuity equation holds true for the marginal
(spatial) probability density $\rho = \int f d^3u $ and  $
\overrightarrow{v} \doteq {\frac{1}\rho } \int \overrightarrow{u}
f d^3u$. That has a devastating effect on the form of the
corresponding momentum conservation law in the large friction
regime.

The associated  Smoluchowski process  with a forward drift
$\overrightarrow{b}(\overrightarrow{x}) =
\frac{\overrightarrow{F}}{m\beta }$
 is analyzed in terms  of the  normalized Wiener process
 $\overrightarrow{W}(t)$:  the infinitesimal increment of the
 configuration (position) random variable $\overrightarrow{X}(t)$ reads:
 $d\overrightarrow{X}\left( t\right) =\frac{\overrightarrow{F}}{\ m\beta }dt+%
\sqrt{2D}d\overrightarrow{W}\left( t\right)$ \,   $\longrightarrow $\,
$\partial _t\rho =
D\triangle \rho -  \overrightarrow{\nabla }\cdot (\rho
\overrightarrow{b})$.\\
In the hydrodynamical picture, we infer  the closed system of two
(special to Markovian diffusions !) local
 conservation laws in the form appropriate  for the Smoluchowski process,
(remember about specific functional forms of  $\Omega $ and $Q$):
\begin{eqnarray}
\partial _{t}\rho + \overrightarrow{\nabla }\cdot \left(
\overrightarrow{v} \rho \right)%
 &=&0 \\
( \partial _{t} + \overrightarrow{v}
\cdot \overrightarrow{\nabla }) \overrightarrow{v}%
 &=& \overrightarrow{\nabla }
\left( \Omega -Q\right)
\end{eqnarray}

\subsection{ Free random dynamics with no friction}

 Now, $\frac{dx}{dt} =u $ and $\frac{du}{dt} =A\left( t\right)$,
hence:
\begin{equation}
\frac{\partial f}{\partial t}  + u\frac{\partial f}{\partial x}  =  C(f) =   q\frac{%
\partial ^{2}f}{\partial u^{2}}
\end{equation}
We know  \cite{czopnik} the transition density:
\begin{equation}
p\left( \left. x,u,t\right| x_{0},u_{0},t_{0}=0\right) =
\frac{1}{2\pi }\frac{%
\sqrt{12}}{2qt^{2}}\exp \left[ -\frac{\left( u-u_{0}\right) ^{2}}{4qt}-\frac{%
3\left( x-x_{0}-\frac{u+u_{0}}{2}t\right) ^{2}}{qt^{3}}\right]
\end{equation}
By choosing    an initial phase space density:
\begin{equation}
f_{0}\left( x,u\right) =\left( \frac{1}{2\pi a^{2}}\right) ^{\frac{1}{2}%
}\exp \left( -\frac{\left( x-x_{ini}\right) ^{2}}{2a^{2}}\right)
\left( \frac{1}{2\pi b^{2}}\right) ^{\frac{1}{2}}\exp \left(
-\frac{\left( u-u_{ini}\right) ^{2}}{2b^{2}}\right)\, .
\end{equation}
so that  $f\left( x,u,t\right) =\int p\left( \left. x,u,t\right|
x_{0},u_{0},t_{0}=0\right) f_{0}\left( x_{0},u_{0}\right)
dx_{0}du_{0}$ and  passing   to the hydrodynamical picture
(unpleasant steps), we observe  that  $\int C(f)du =0$
and $\int u C(f) du =0$ which yields  the following outcomes,
\cite{czopnik}:
\begin{equation}
\rho \left( x,t\right) = \left(
\frac{1}{2\pi \left( a^{2}+b^{2}t^{2}+\frac{2}{3}qt^{3}\right)
}\right) ^{\frac{1}{2}}\exp \left(
-\frac{\left( x-x_{ini}-u_{ini}t\right) ^{2}}{2\left( a^{2}+b^{2}t^{2}+\frac{%
2}{3}qt^{3}\right) }\right)
\end{equation}
\begin{equation}
\rho \left( u,t\right) = \left(
\frac{1}{2\pi \left( b^{2}+2qt\right) }\right) ^{\frac{1}{2}}\exp
\left( -\frac{\left( u-u_{ini}\right) ^{2}}{2\left(
b^{2}+2qt\right) }\right)
\end{equation}
\begin{equation}
\left\langle u\right\rangle _{x}=
u_{ini}+\frac{b^{2}t+qt^{2}}{a^{2}+b^{2}t^{2}+\frac{2}{3}qt^{3}}\left[
x-x_{ini}-u_{ini}t\right] \doteq v
\end{equation}
\begin{equation}
\left\langle u^{2}\right\rangle _{x}-\left\langle u\right\rangle
_{x}^{2}=\frac{q\,t^{3}\,\left(
2\,b^{2}+q\,t\right) +3\,a^{2}\,\left( b^{2}+2\,q\,t\right) }{%
3\,a^{2}+t^{2}\,\left( 3\,b^{2}+2\,q\,t\right) } \doteq \frac{P_{kin}}\rho
\end{equation}
This  implies the local momentum  conservation law:
\begin{equation}
\left( \frac{\partial }{\partial t}+ v\cdot \nabla \right) v= -\frac{\nabla
P_{kin}}{\rho }= + 2\left( d^{2}\right) \nabla \left[ \frac{\Delta
\rho ^{1/2}}{\rho ^{1/2}}\right] \doteq + \nabla Q
\end{equation}
with
\begin{equation}
d^{2}(t)=a^{2}b^{2}+2a^{2}qt+\frac{2}{3}b^{2}qt^{3}+
\frac{1}{3}q^{2}t^{4} \doteq D^2(t)  \, .
\end{equation}
Remember about: $d^{2}\left( t\right) \doteq \left( \left\langle
u^{2}\right\rangle _{x}-\left\langle u\right\rangle
_{x}^{2}\right) \left( \left\langle x^{2}\right\rangle
-\left\langle x\right\rangle ^{2}\right)$ and notice that
$d^2(t)\geq a^2b^2$.

\subsection{Noiseless limit, $C(f)= 0$ for all $f$}

Upon  disregarding  random forcing (set $q\rightarrow 0$ in Eq. (19)), we
arrive at:
\begin{equation}
\frac{\partial f}{\partial t}+u\frac{\partial f}{\partial x}+\frac{F}{m }
\frac{\partial f}{\partial u}= 0
\end{equation}
where clearly  $ \int C(f)du =0 = \int u C(f)du = \int u^2 C(f)du$.

Things now  look classical and there is good reason for that, since Eq. (28) is
the familiar Liouville equation. However this "classical look" appears slightly
deceiving.

Indeed,  the $q \rightarrow 0$ limit of the frictionless free dynamics gives rise to:
\begin{equation}
f\left( x,u,t\right) =\frac{1}{2\pi
\sqrt{a^{2}b^{2}}}\exp \left( -\frac{\left( u-u_{ini}\right)
^{2}}{2b^{2}}-\frac{\left( x-x_{ini}-tu\right)
^{2}}{2a^{2}}\right) \, .
\end{equation}
with marginals:
\begin{equation}
{\rho }\left( u,t\right) =\left( \frac{1}{2\pi b^{2}}\right) ^{%
\frac{1}{2}}\exp \left( -\frac{\left( u-u_{ini}\right)
^{2}}{2b^{2}}\right)
\end{equation}
and
\begin{equation}
{\rho }\left( x,t\right) =\left( \frac{1}{2\pi \left(
a^{2}+b^{2}t^{2}\right) }\right) ^{\frac{1}{2}}\exp \left(
-\frac{\left( x-x_{ini}-u_{ini}t\right) ^{2}}{2\left(
a^{2}+b^{2}t^{2}\right) }\right) \, .
\end{equation}
The local moments read:
\begin{equation}
\left\langle u\right\rangle _{x}=u_{ini}+\frac{b^{2}t}{a^{2}+b^{2}t^{2}}%
\left( x-x_{ini}-u_{ini}t\right)
\end{equation}
and
\begin{equation}
\left\langle u^{2}\right\rangle _{x} - \left\langle u\right\rangle _{x}^{2}=
\frac{a^{2}b^{2}}{a^{2}+b^{2}t^{2}}
\end{equation}
which yields  the (local)  momentum conservation law  in the fairly nonclassical form:
\begin{equation}
\left( \frac{\partial }{\partial t}+\left\langle u\right\rangle
_{x}\nabla
\right) \left\langle u\right\rangle _{x}=2a^{2}b^{2}\nabla \left[ \frac{%
\Delta {\rho }\left( x,t\right) ^{1/2}}{{\rho
}\left( x,t\right) ^{1/2}}\right] \doteq  \nabla Q
\end{equation}
By setting $a\cdot b= \frac{\hbar }{2m}$ we recover the standard
 quantum mechanical "hydrodynamics", Eq. (7), to be compared with the Brownian variant
of the law, Eq. (8).
 Notice that $<x^2> - <x>^2 = a^2 + b^2t^2$ and:
\begin{equation}
 a^2b^2 =  \left( \left\langle u^{2}\right\rangle
_{x}-\left\langle u\right\rangle _{x}^{2}\right) \left( \left\langle
x^{2}\right\rangle -\left\langle x\right\rangle ^{2}\right)= {\frac{\hbar ^2}{4m^2}}
\end{equation}
for all times. That is another  expresssion for  the standard quantum
mechanical  Heisenberg indeterminacy relation, see e.g.
\cite{falco}-\cite{hall}.

We recall  that  $Q_q= - {\frac{\hbar
^2}{2m^2}}{\frac{\Delta \rho ^{1/2}}{\rho ^{1/2}}}$  is the
  de Broglie - Bohm quantum potential. Is there anything specific or mysterious in its
origin  and physical meaning?

\section{Miscellaneous contexts:  Hamilton-Jacobi, Liouville,
Kramers equations, calculus of variations}

Let us recall  so-called  \it wave equations \rm  of   classical mechanics:
\begin{equation}
\partial _t\rho = - \nabla \cdot (\rho \frac{\nabla S}m)
\end{equation}
with
\begin{equation}
\partial _t S + \frac{(\nabla S)^2}{2m} + V = 0
\end{equation}
Rename: $\frac{S}m \rightarrow S$, set $v = \nabla S$, eventually
take a gradient of the above Hamilton-Jacobi  equation. Then, we have:
\begin{equation}
\partial _t\rho = - \nabla \cdot (v\rho )
\end{equation}
and
\begin{equation}
\partial _t v + (v\cdot \nabla )v = - \nabla V
\end{equation}

Clearly, in the above there is \it nothing alike the  \rm  $\nabla Q$
contribution, so characteristic to our  previous  dynamical examples,  cf.
Eqs. (7)-(9), (11), (12), (18), (26).
 What  is the  primary reason of so  conspicuous absence of that term in Eq. (39) ?

To set a connection with the  Liouville equation  we follow a standard assumption,
\cite{holland}:
  assign a \it  unique \rm  momentum value
at each space point and consider phase-space densities as generalized functions
\begin{equation}
f_0(x,p) = \rho _0(x) \delta (p - \nabla S_0(x))\,  \,  \longrightarrow \, \,
f(x,p.t) = \rho (x,t) \delta (p - \nabla S(x,t))
\end{equation}
which (weakly) solve
\begin{equation}
\frac{\partial f}{\partial t}+u\frac{\partial f}{\partial x}+\frac{F}{m }
\frac{\partial f}{\partial u}= 0 \, .
\end{equation}

In view of the fact that the Liouville equation  preserves in time  the
\it  precise \rm  knowledge of initial data, we have:
\begin{equation}
f_0(x,p) = \delta (x-x_0) \delta (p- p_0) \rightarrow  f(x,p,t) =
\delta (x- x(t,x_0,p_0))\delta (p - p(t,x_0,p_0))
\end{equation}
to be compared with the  density  function, Eq.  (29).

As a side remark, let us notice that the Hamilton-Jacobi  equation can be derived
 via the  least action principle by employing  the Lagrangian  density
\begin{equation}
{\cal{L}} =  \rho\left[ \partial _t S   +  {\frac{1}2} (\nabla S)^2 + \frac{V}m \right]
\end{equation}
with $ \rho $ and $S$ considered as canonically conjugate variables, \cite{holland}.

 Where has gone  our   $\nabla Q$  (and $Q$ itself)  ?

Let us come back to the previous  $q\rightarrow 0$ free motion  case, Eqs. (21) and (29).
For  all times $t\geq 0$  both  spatial and  velocity  parts
of the  phase-space  density are  well behaved functions (not  Dirac deltas).
Hence, and indispensable, crucial step has been there to  admit from the beginning both
 the spatial and  momentum (velocity) indeterminacy (spreading, unsharpness).
At time $t=0$, we assign to each point $x$ a "bunch" of possible (to be picked
up at random from a given probability law) momenta - a Gaussian distribution
of momenta at each  spatial point and in addition we adopt  a definite
 probability law for the position variable.
As an  immediate outcome, we get Eq.(34) i.e.:
\begin{equation}
\left( \frac{\partial }{\partial t}+ v \cdot \nabla
\right) v =  \nabla Q
\end{equation}
where: $+ 2a^{2}b^{2}\left[ \frac{\Delta {\rho }\left( x,t\right) ^{1/2}}{{\rho
}\left( x,t\right) ^{1/2}}\right] \doteq   Q$.  This is the  gradient form of:
\begin{equation}
\partial _tS +\frac{1}2 (\nabla S)^2 -  Q = 0
\end{equation}
which derives (via the standard variational calculus) from the Lagrangian density:
\begin{equation}
{\cal{L}} = \rho\left[ \partial _t S   +  {\frac{1}2} (\nabla S)^2  + {\frac{u^2}2} \right]
\end{equation}
to be compared with the previous "precise" (sharp)  momentum variant in the
absence  of conservative forces:
\begin{equation}
{\cal{L}} = \rho\left[ \partial _t S   +  {\frac{1}2} (\nabla S)^2    \right] \, .
\end{equation}

 Now we need  to  come back to Eqs. (1)-(5), where entangled relationships
 among $Q$, Fisher information and local (information)  entropy production
 were established for the Brownian motion.

In direct affinity  with Eq. (46), we can develop  the Hamiltonian
formalism ($\rho$ and $S$ are canonically conjugate, $D=a b$), \cite{reginatto}
which  employs :
\begin{equation}
H = \int {\cal{H}} dx = \int dx\, \rho \cdot \left[ {\frac{1}2} (\nabla S)^2  +
{\frac{u^2}2} \right] = \int dx \cdot \rho {\frac{1}2} (\nabla S)^2   + 2D^2\cdot  F_X \, .
\end{equation}

We can now devise convincing arguments which  relate the emergence of $F_X$ and $Q$ in the
above with the a priori introduced  and \it  simultaneously valid  \rm  spatial and
 momentum indeterminacy.
To this end we shall discuss the position and velocity unsharpness issue from the  two, looking
diverse, perspectives.

Concerning the spatial (position) indeterminacy, we  need the existence
of the $F_X$ term in Eq. (48). From the properties of the Fisher information,
\cite{frieden,hall}, there   follows that $F_X$ goes to infinity when the spatial
probability density approaches the delta function (sharp localization) limit.
(The same happens when the probability density is discontinuous or vanishes
over certain interval). Hence, the Hamiltonian (48) is properly  defined only
in case of the nonsingular,  unsharp spatial localization.

With regard to the velocity (momentum) unsharpness, let us invoke classic
observations in the  so-called  quantum theory of motion
(Bohm theory, Holland (1993)),  where one  argues  as
follows (notice a "subtle" difference if  compared to our probabilistic
arguments).

Equations: $\partial _t\rho = - \nabla (v\rho )$ where $v = \frac{1}m \nabla S$ and
\begin{equation}
\left( \partial _t + {\frac{1}m} v\cdot \nabla \right) v =  - \nabla (V + Q_q)
\end{equation}
where $Q_q \simeq  - Q$ \it imply \rm  that the distribution function:
\begin{equation}
f(x,p,t) = \rho (x,t) \delta \left[ p - \nabla S(x,t)\right]
\end{equation}
obeys the law of evolution:
\begin{equation}
\partial _t f + \frac{p}m \cdot \nabla _x f + \nabla _x(V + Q_q )\cdot
\nabla _p f =0\, .
\end{equation}

This equation reduces to the  classical  Liouville one  only  when $Q_q =0$,
while the whole body of our previous discussion  has  explicitly referred  to
the  Liouville equation as a primary building block of the theory.
Consequently, in this  context, a possibility that $p$
(respectively $u$) can be sharply defined at each spatial
point is definitely excluded, even if the spatial contribution is
a priori assumed to be  unsharp.

{\bf Comment:}
For comparison with  the previous  random motion discussion, one should
realize that the continuity (and thus Fokker-Planck) equation plus the
Hamilton-Jacobi type equation of the  general form (we formally use
$V/m $ instead of  more correct $\Omega $):
\begin{equation}
\partial _tS +\frac{1}2 (\nabla S)^2 \pm (\Omega -  Q) = 0
\end{equation}
referring  to the local conservation law:
\begin{equation}
\left( \frac{\partial }{\partial t}+ v \cdot \nabla
\right) v = \mp \nabla (\Omega - Q)
\end{equation}
both  derive via the calculus of variations from,
\cite{skorobogatov,reginatto1}:
\begin{equation}
L = \int {\cal{L}} dx = \int \rho\left[ \partial _t S   +  {\frac{1}2} (\nabla S)^2  \pm
\left({\frac{u^2}2} + \Omega \right) \right] dx
\end{equation}
The related  Hamiltonian reads:
\begin{equation}
H = \int {\cal{H}} dx = \int dx\, \rho \cdot \left[ {\frac{1}2} (\nabla S)^2  \pm
\left( {\frac{u^2}2} + \Omega \right ) \right]
\end{equation}
All that refers exclusively to  the general phase-space equation:
\begin{equation}
\frac{\partial f}{\partial t}+u\frac{\partial f}{\partial x}+\frac{%
{F}}{m }\frac{\partial f}{\partial u}=C\left( f\right)
\end{equation}
and \it not  \rm to any  equation of the form (51).

\section{Supplement: variational arguments in the theory of the Brownian motion}

In connection with formulas (1)-(5) it is instructive to recall  that in the
Lagrangian formulation of the theory of random motion \cite{santos}  the maximum
rate of the information entropy increase has been found to maximize  the
Fisher information. After suitable  notation adjustments, we  realize that
in Ref. \cite{santos} it is exactly $- F_X$ that
is minimized to yield the free spatial Brownian motion.

In the very same case, \cite{guth}, the relationship (cf. Eq. (4) for comparison):
\begin{equation}
<X^2> <u(X,t)^2> = D
\end{equation}
was established for the heat kernel solution of $\partial _t\rho  =
D\Delta \rho $, provided there holds  $<X>=0=<u(X,t)>$.
For more general solutions of the heat equation for which the mean values of
$X(t)$ and $u(X(t),t)$ do vanish, one arrives at  a general indeterminacy
relationship with an obvious affinity to the previously mentioned
 Cramer-Rao inequality:
\begin{equation}
<X^2> <u(X,t)^2> =\left[ D^2 \int dx \, {\frac{1}\rho }\cdot
(\nabla \rho )^2 \right]^{1/2}\cdot \left[ \int dx \,  x^2\cdot
  \rho \right]^{1/2}
  \geq  D  \, .
\end{equation}

Quite in parallel, a bit more general case was addressed in Ref. \cite{viola},
where a general (nonvanishing forward drift)  one dimensional diffusion
process with  (time-dependent)
diffusion coefficient $D(t)$ was considered,  under slightly weaker
restrictions:  $<u(X,t)> = 0$ while $<X> \neq 0$.
The problem addressed, has been  an issue of when  the product of variances
$[<X^2> - <X>^2]\cdot [<u(X,t)^2>]$ is minimized.

The outcome is that a minimum is reached for a concrete product value
equal $D^2(t)$ and that a necessary and sufficient condition for the probability
density $\rho (x,t)$ to yield that minimum, is that  it has a Gaussian form:
\begin{equation}
\rho (x,t) = {\frac{1}{(2\pi )^{1/2} [<X^2> - <X>^2]^{1/2}}}\cdot
\exp \left[ - {\frac{(x - <X>)^2}{2[<X^2> - <X>^2]}} \right]
\end{equation}
in agreement with our previous discussion. For nongaussian probability densities,
an inequality of the type (58) holds true.

Indeed, it is a classic observation, \cite{cramer}, that for a generic
probability density  $\rho _{\alpha }(x,t)$ with the first moment
$\int x\cdot \rho (x,t) dx = f(\alpha ,t)$
and finite second moment, for which there exist both  partial derivatives
$\frac {\partial \rho _{\alpha }(x,t)} {d \alpha }$ and
$\frac{\partial f(\alpha ,t)} {\partial \alpha }$ (for all $\alpha $ in  an
interval in $R^1$ or generally  in $R^1$,  and for almost all $x \in R^1$),
then we arrive at an inequality:
\begin{equation}
\int (x- \alpha )^2 \rho _{\alpha }(x,t) dx \cdot \int \left(
 {\frac{\partial ln \rho _{\alpha }}{\partial \alpha }} \right)^2 \rho _{\alpha }(x,t) dx
 \geq \left( \frac{df(\alpha ,t)}{d\alpha }\right)^2 \, .
 \end{equation}


\begin{thebibliography}{99}
\bibitem{cohen} E. G. D. Cohen, L. Rondoni, Physica {\bf A 306},
117, (2002)
\bibitem{rondoni} L. Rondoni, E. G. D. Cohen, Physica {\bf D
168-169}, 341, (2002)
\bibitem{gaspard} P. Gaspard, "Chaos, Scattering and Statistical
Mechanics", (Cambridge University Press, Cambridge, 1998)
\bibitem{gaspard1} P. Gaspard, J. Stat. Phys. {\bf 88}, 1215,
(1997)
\bibitem{gaspard2} P. Gaspard, G. Nicolis, J. R. Dorfman,
"Diffusive Lorentz gases and multibaker maps are compatible with
irreversible thermodynamics", Los Alamos arXiv:nlin.CD/0210060,
(2002)
\bibitem{gar} P. Garbaczewski, Phys. Rev. {\bf E 59}, 1498, (1999)
\bibitem{gar1} P. Garbaczewski, Physica {\bf A 285}, 187, (2000)
\bibitem{gar2}  R. Czopnik, P. Garbaczewski, Phys. Rev. {\bf E 63},
021105, (2001)
\bibitem{czopnik} R. Czopnik, P. Garbaczewski, Physica {\bf A}, (2002), in press;
Los Alamos arXive:cond-mat/0202463, (2002)
\bibitem{frieden} B. Roy Frieden, Phys. Rev. {\bf A 41}, 4265, (1990)
\bibitem{luo} S. Luo, J. Phys. A: Math. Gen. {\bf 35}, 5181, (2002)
\bibitem{carlen} E. A. Carlen, J. Funct. Anal. {\bf 101}, 194, (1991)
\bibitem{rao} C. R. Rao, Bull. Calcutta Math. Soc. {\bf 37}, 81, (1945),
\bibitem{cramer} H. Cramer, "Mathematical Methods of Statistics",
(Princeton University Press, Princeton, 1946)
\bibitem{holland}   P. R. Holland, "Quantum Theory of Motion",
(Cambridge University Press, Cambridge, 1993)
\bibitem{falco} D. de Falco et al, Phys. Rev . Lett. {\bf 49}, 181, (1982)
\bibitem{golin} S. Golin, J. Math. Phys., {\bf 26}, 2781, (1985)
\bibitem{viola} F. Illuminati, L. Viola, J. Phys. A: Math. Gen. {\bf 28},
2953, (1995)
\bibitem{hall} M. J. W. Hall, Phys. Rev. {\bf A 64}, 052103, (2001)
\bibitem{reginatto} M. J. W. Hall, M. Reginatto, J. Phys. A: Math. Gen.
{\bf 35}, 3289, (2002)
\bibitem{skorobogatov}  G. A. Skorobogatov, Rus. J. Phys. Chem., {\bf 61},
509, (1987)
\bibitem{reginatto1} M. Reginatto, Phys. Rev. {\bf A 58}, 1775, (1998)
\bibitem{santos} E. Santos, Nuovo Cim. {\bf 59 B}, 65, (1969)
\bibitem{guth} E. Guth, Phys. Rev. {\bf 126}, 1213, (1962)
\end{thebibliography}
\end{document}